\documentclass[%
 reprint,
superscriptaddress,
groupedaddress,
nofootinbib,
 amsmath,amssymb,
 aps,
 prl,
]{revtex4-2}

\usepackage{graphicx}% Include figure files
\usepackage{dcolumn}% Align table columns on decimal point
\usepackage{bm}% bold math
\usepackage{hyperref}% add hypertext capabilities

\usepackage{xcolor}

\usepackage{cancel}
\usepackage[normalem]{ulem}
\usepackage{hyperref}

\usepackage{amsmath}

\usepackage{listings}

\definecolor{codegreen}{rgb}{0,0.6,0}
\definecolor{codegray}{rgb}{0.5,0.5,0.5}
\definecolor{codepurple}{rgb}{0.58,0,0.82}
\definecolor{backcolour}{rgb}{0.95,0.95,0.92}

\lstdefinestyle{mystyle}{
    backgroundcolor=\color{backcolour},   
    commentstyle=\color{codegreen},
    keywordstyle=\color{magenta},
    numberstyle=\tiny\color{codegray},
    stringstyle=\color{codepurple},
    basicstyle=\ttfamily\footnotesize,
    breakatwhitespace=false,         
    breaklines=true,                 
    captionpos=b,                    
    keepspaces=true,                 
    numbers=left,                    
    numbersep=5pt,                  
    showspaces=false,                
    showstringspaces=false,
    showtabs=false,                  
    tabsize=2
}

\usepackage{rotating}
\usepackage{cases}

\begin{document}

\preprint{APS/123-QED}

\title{A trade-off between hydrodynamic performance and morphological bias limits the evolution of symmetric lattice animal wings}

\author{Sif F. Arnbjerg-Nielsen$^1$}
\author{Matthew D. Biviano$^1$}
\author{Annette Cazaubiel$^2$}
\author{{Alexander Schødt}$^1$}
\author{Andreas Carlson$^2$}
\author{Kaare H. Jensen$^1$}
\email{khjensen@fysik.dtu.dk}
\affiliation{$^1$Department of Physics, Technical University of Denmark, DK-2800 Kgs. Lyngby, Denmark\\
$^2$Department of Mathematics, University of Oslo, 0315 Oslo, Norway}

\date{\today}% It is always \today, today,

\begin{abstract}
{Bristled and membranous insect wings have co-evolved despite apparently serving the same functionality. We emulate flight physics using an automated free-fall experiment to better understand how and why several distinct wing forms may have developed. Biomimetic two-dimensional lattice animals were laser cut from a continuous sheet of paper, and their descent in a settling tank was tracked using a camera.}
Data from {31} generic symmetric polyominos ({1,692} experiments) reveal that morphology impacts the drag coefficient $C_D$ and hence flight efficiency.  Some polyominos rapidly sediment while others remain suspended for longer. Positioning the search for an optimal shape within an evolutionary context, we relinquished control of the automated setup to a genetic algorithm. Hereditary information passes between generations in proportion to fitness and is augmented by rare mutations. High-performing morphologies are observed, but {experimental repeats do} not re-evolve the same shapes. Adaptation rates also differ if the selective pressure is reversed from suspension to sedimentation. Our data ({50,086} experiments) reveal several physical sources of indeterminism in the simulated natural selection process, including fluid flow variability and morphological entropy. This provides experimental support for the idea that the lack of convergence in insect wing shape may partly reflect the competition between fitness and entropy, making it nearly impossible to achieve unique optimal forms.
\end{abstract}
%1,746, 32

\maketitle
\section{Introduction}

{Natural selection drives the evolution of form} and function of living organisms to fit their environment constrained by cost \cite{lauder1981form}. Unveiling the sources of predictability, or the lack thereof, in evolution is thus a fundamental biophysical problem. Sometimes, populations can respond as expected to ecological changes or challenges in a process known as convergent evolution \cite{stern2013genetic}. Other times, {emerging traits appear to be non-deterministic, but physically well-defined examples of this are rare}. 

\begin{figure*}[ht!]
    \centering
    \includegraphics[width = 18cm]{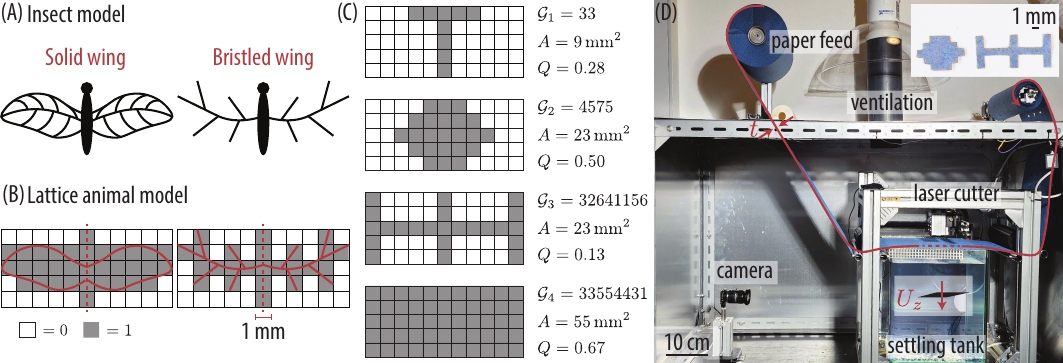}
    \caption{Lattice animals and experimental setup. (A) Natural wing morphologies include solid and bristled {shapes}. (B) Wings are represented by $5\times 11$ matrix with entries $1$ ({solid}) and $0$ (void) based on a 1x1 mm$^2$ unit cell grid.  We only consider symmetric and simply rook-wise connected morphologies \cite{read1962contributions,golomb1996polyominoes}. (C) Examples of lattice animals of area $A$, compactness $Q$ (Eq. \eqref{eq:Q}), and genome $\mathcal{G}$ (see details in the text). (D) Experimental setup used to quantify the vertical terminal velocity $U_z$ from video recordings of the lattice animal. The polyominos {(insert)} are cut by a laser from a continuous paper sheet and fall {into} the settling tank. %{Only the last $\approx$ 13 cm of the fall is recorded ($\sim25\sqrt{A}$) to ensure terminal velocity is reached.}
    }
    \label{fig:fig1}
\end{figure*}

Insect wing morphology may serve as a basic example of evolutionary variability{, where} distinct wing geometries can be observed in extant species occupying similar environments \cite{salcedo2019computational,sun2023colloquium,engels2021flight, engels2021flight,luna2023fluid,o2022efficiency,polilov2015small}. A specific example is the case of porous (bristled) and solid (membranous) wings in small flying insects (Fig \ref{fig:fig1}A) \cite{engels2021flight,ford2019aerodynamic,o2022efficiency}. 
Most insect wings are solid and impermeable to air \cite{wootton1992functional}. During flight, fluid flows around the cuticular membrane strengthened by longitudinal and transverse veins. In contrast, some air can leak through bristled wings, comprising {rake}-like appendages attached to a slender rod {\cite{engels2021flight,sun2023colloquium,lee2020optimal}}.
%{While a diminished body size is associated with a range of adaptions \cite{polilov2015small}, bristles have largely been attributed to the altered aerodynamics during locomotion \cite{engels2021flight,luna2023fluid,o2022efficiency}.} 
However, the quantitative reason why bristled and membraneous insect wings co-exist remains unclear.
%Most insect wings are solid and impermeable to air \cite{wootton1992functional}. During flight, fluid flows around the cuticular membrane strengthened by longitudinal and transverse veins. Other wing designs, however, exist. An example is the bristled wing, comprising {rake}-like appendages attached to a slender rod {\cite{engels2021flight,sun2023colloquium,lee2020optimal}} (Fig \ref{fig:fig1}A). 

Recent work (see e.g., \citet{lee2017aerodynamics}) has demonstrated that the leakiness of bristled wings reduces its aerodynamic performance compared to the solid wing {\cite{lee2017aerodynamics}}. However, the bristled wing is also lighter. Per unit mass, permeable wings can, therefore, sometimes yield an evolutionary advantage {\cite{polilov2015small,luna2023fluid,o2022efficiency,engels2021flight,kolomenskiy2020aerodynamic,sun2023colloquium,engels2021flight,lee2017aerodynamics,ford2019aerodynamic}}. To our knowledge, however, the precise balance between performance and shape, has not been explored in an evolutionary and statistical context. 

This letter positions morphology and performance in a {synthetic} evolutionary framework. Artificial wings settling under gravity provide a {model of parachuting flight \cite{lee2020stabilized}}, while evolution is mimicked by a genetic algorithm \cite{ramananarivo2019improving,eloy2013best,spall2005introduction,ramananarivo2019improving}. We introduce polyomino wings to avoid the limitations associated with pre-defined morphologies \cite{luna2023fluid,lee2017aerodynamics,lee2020optimal}. We begin by introducing the experimental methodology. Subsequent sections lay out the results. Finally, the the implications for shape evolution are discussed. 

\section{Methods}
\subsection{Morphology}
To probe the relation between morphology and fitness, we begin by introducing a two-dimensional insect model. {While insect wings are not fully flat \cite{schubnel2023flat}, their membranous nature allow this approximation \cite{salcedo2019computational}.} To model wing shape, we adopt a lattice animal representation which permits us to probe biologically relevant morphologies while retaining nearly complete geometric freedom (Fig. \ref{fig:fig1}B-C).  Each lattice animal is represented by a $5 \times 11$ matrix. Every matrix entry corresponds to a (physical) $1\times 1$ mm$^2$ square represented by two possible values: $1$ (solid) and $0$ (void). The insect body (6th matrix column) is always solid, while the wing morphology (columns 1-5 and 7-11) can vary. For simplicity, we restrict our attention to symmetric and simply rook-wise connected wings \cite{read1962contributions,golomb1996polyominoes}. An example of a lattice animal matrix is:
\begin{equation}
\begin{array}{cccccccccccc}
\cline{6-8}
&0 & 0 & 0 & 0 & \multicolumn{1}{|c}{1} & \color{blue}{1} & \multicolumn{1}{c|}{\color{gray}{1}} &  \color{gray}0 &  \color{gray}0 &  \color{gray}0 &  \color{gray}0 \\\cline{5-5}\cline{9-9}
&0 & 0 & 0 & \multicolumn{1}{|c}{1} & 1 & \color{blue}{1}  & \color{gray}{1}  & \multicolumn{1}{c|}{\color{gray}{1}} &  \color{gray}0 &  \color{gray}0 &  \color{gray}0 \\\cline{4-4}\cline{10-10}
M=&0 & 0 & \multicolumn{1}{|c}{1} & 1 & 1 & \color{blue}{1}  & \color{gray}{1}  & \color{gray}{1}  & \multicolumn{1}{c|}{\color{gray}{1}}  &  \color{gray}0 &  \color{gray}0 \\\cline{4-4}\cline{10-10}
&0 & 0 & 0 & \multicolumn{1}{|c}{1} & 1 & \color{blue}{1}  & \color{gray}{1}  & \multicolumn{1}{c|}{\color{gray}{1}}  &  \color{gray}0 & \color{gray}0 &  \color{gray}0 \\\cline{5-5}\cline{9-9}
&0 & 0 & 0 & 0 & \multicolumn{1}{|c}{1} & \color{blue}{1}  & \multicolumn{1}{c|}{\color{gray}{1}}  &  \color{gray}0 &  \color{gray}0 &  \color{gray}0 &  \color{gray}0 \\\cline{6-8}
&\multicolumn{5}{c}{\underbrace{\hspace{5em}}_{\text{left wing}}} & \multicolumn{1}{c}{\underbrace{\hspace{0em}}_{\text{body}}}& \multicolumn{5}{c}{ \color{gray}\underbrace{\hspace{5em}}_{\text{ \color{gray}right wing}}}\end{array}
\nonumber
\end{equation}
The insect outline is formed by tracing the matrix's value-$1$ entries, which, in this case forms an approximately \emph{oval} wing (Fig. \ref{fig:fig1}C, S{2}). 
 
Each polyomino is {uniquely identified} by its genome, found by concatenating the  elements of the first $5$ vertical matrix columns starting in the lower left corner. The rest are redundant due to symmetry. For the matrix given above, it is $\mathcal{G}=00000\,00000\, 00100\,01110\,11111 (2)$, which we can think of as a binary ($\text{base}\;2)$ number. This notation provides a convenient tool to track and compare different shapes. In the more compact decimal (base 10) representation, it is $\mathcal G =4575\,(10)$ \footnote{Note that morphologically similar forms are not necessarily numerically close in the decimal representation of the genome. However, a numerically large genome indicates a large area or a sparse morphology, while a small genome indicates a small area or a compact shape.}. For a genome of length $N$ with $n$ unit entries, there are $W=N!/[(N-n)!n!]$ possible combinations. For instance, we can identify $W\approx 2\times 10^6$ potential morphologies of area $A=23$ (genomes with $n=9$ unit entries among $N=5\times 5=25$ places). However, because we restrict ourselves to simply rook-wise connected wings, the accessible number is significantly smaller: A brute-force search for morphologies of area $23$ revealed just $7325$ unique polyominoes with up-down symmetry accounted for (Fig. \ref{fig:uniqueInsects}). 

\subsection{Settling experiment}
To quantify the basic hydrodynamic properties of each two-dimensional shape (or corresponding {genome}), we use free-fall experiments \cite{list1971free,jayaweera1972equivalent,bagheri2016drag}. Facilitated by advances in manufacturing technology and robotics (see, e.g., \citet{howison2020large}), we utilize an Automated Settling APparatus henceforth abbreviated as ASAP. The first use of this device was previously reported by members of our team: \citet{2409.05514}. Briefly, the lattice animal was cut by a laser (laser: N40630, 2D-{plotter}: Neje 3 Laser Engraver and Cutter, Neje, China) from a continuous paper roll (blue 200m x 210mm single-ply, Katrin, Finland) of density $\rho_p\approx 1115$ kg/m$^3$ (wet) and thickness, $t\approx0.1$ mm. The shape then fell into the settling tank ({25x25x30 cm}, Fig. \ref{fig:fig1}D) and was tracked by a camera (camera: {Raspberri Pi HQ camera}, Raspberry Pi Foundation, UK, lens: C11-1220-12M-P f12mm, Basler, Germany). Finally, the terminal velocity, $U_z$, was computed by fitting a straight line to the vertical position/time coordinates. In this paper we will use speed and velocity interchangeably to describe the vertical descent rate. Each experiment took approximately one minute. A total of {51,778}  experiments are included in this study. 

To determine the mean vertical settling speed $U_z$ and the standard deviation $\sigma_{U_z}$, the sedimentation experiments were repeated for each shape $21-59$ times (Fig. \ref{fig:fig2}A).
%leaving 21-59 recorded falls after filtering.} 
On average, the relative error was $\sigma_{U_z}/U_z\approx 0.11$. We interpret this as a result of the well-established sensitivity to initial and environmental conditions in free-fall experiments \cite{esteban2019study,tinklenberg2023thin,howison2020large} (see also Fig \ref{fig:backgroundVariation}). Data from a limited number of failed experiments were rejected based on Chauvenet's criterion \cite{taylor1997introduction}.

Shapes of area $A\approx 20$ mm$^2$ typically fell at a speed of $U_z \approx 1$ cm/s, corresponding to the Reynolds number range 
\begin{equation}
Re = U_z d_{\text{eq}}/\nu\approx {30}-110,\label{eq:Re}
\end{equation} where $d_{\text{eq}}=2\sqrt{A/\pi}$ is the diameter of an equal-area disk $\nu=10^{-6}$ m$^2$/s is the kinematic viscosity of the fluid (water). During settling, the shapes' surface normal remain parallel to the direction of gravity, and the out-of-plane bending was negligible. Rotation was observed during settling, but the shapes never flipped. {Detailed flow characterization was carried out with flowtrace for ImageJ \cite{gilpin2017flowtrace} (Image aqcuisition: FlowMaster 4D-PTV / Shake-the-Box, particles: polyamide 60 \textmu m, both from LaVision GmbH, Germany).} Additional details of the experimental setup are provided in the supplementary materials.

\subsection{Drag coefficient $C_D$ and compactness $Q$}
To gauge the relative performance of each lattice animal, we study how the drag coefficient $C_D$ varies with the wing shape and the physical parameters in the problem. We first proceed by describing how drag $C_D$ was obtained. Subsequently, we introduce an morphological metric (the compactness $Q$) to unpack the effects of shape.

Determining the drag coefficient from our experiments is straightforward: At terminal velocity, the drag force $F_D$ equals the effective gravitational force $F_g$ \cite{walker2021estimation,esteban2019study}. The drag force is $F_D=AC_D\rho_fU_z^2/2$ where $\rho_f$ is the fluid density and $C_D$ is the drag coefficient \cite{walker2021estimation,zastawny2012derivation}. Similarly, the gravitational force is $F_g =  \Delta\rho gAt$, where $g$ is the standard gravity and $\Delta\rho = \rho_p-\rho_f$, is the density difference between paper ($\rho_p$) and fluid ($\rho_f$). Equating drag and gravitational forces leads to an expression for the drag coefficient:
\begin{equation}
	C_D= 2\frac{\Delta \rho}{\rho_f}\frac{gt}{U_z^2}. \label{eq:drag}
\end{equation}
Typical measured values of $C_D\approx 1.5$, consistent with previous studies \cite{willmarth1964steady} (Fig. \ref{fig:S1}).

\begin{figure}[t!]
    \centering
    \includegraphics[width = 8cm]{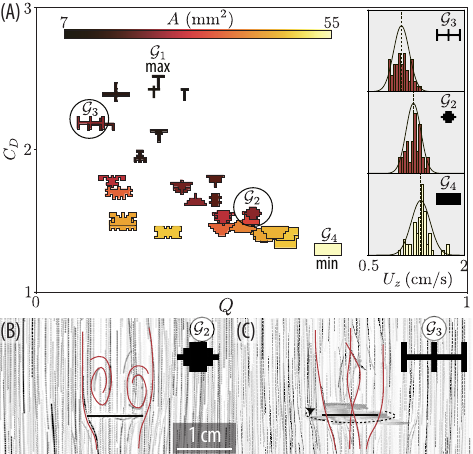}
    \caption{The motion of freely falling objects depends on their size and shape. (A) The drag coefficient $C_D$ (Eq. \ref{eq:drag}) plotted as a function of the compactness $Q$ (Eq. \ref{eq:Q}). Colored shading indicates the polyomino area $A$. Drag decreases with increasing compactness (Fig. \ref{fig:S4}). The maximum drag coefficient observed was that of a small {T-shaped polyomino ($A= 9$ mm$^2$, $C_D=2.5\pm0.08$, genome $\mathcal{G}_1=00000\,\ldots\,00001 \,00001 (2)$)}, while the minimum $C_D$ was observed for a large rectangular wing ($C_D=1.30\pm0.04$, $A=55$ mm$^2$, genome $\mathcal{G}=11111\,\ldots\, 11111 (2)$). The Reynolds number varied from $Re\approx {30}$ ($\mathcal{G}_1$) to $Re\approx 110$ ($\mathcal{G}_4$). Insets: characteristic terminal velocity $U_z$ histograms obtained from {$\approx 55$} repeats. {(B)-(C) Particle trajectories obtained from image data using flowtrace for ImageJ \cite{gilpin2017flowtrace} averaged over 100 frames (Video S1)}. (B) {Separated vortices follow in the wake of a compact polyomino $\mathcal{G}_2$ during settling.}
    (C) Spanwise flow at the shape edge during rotation is observed for a bristled polyomino $\mathcal{G}_3$. Streamlines are drawn to guide the eye. The black horizontal line indicates the position of the falling object.}
    \label{fig:fig2}
\end{figure}

Recall that even for a moderately size lattice animal (say, area $A=23$ mm$^2$) there are at least $7325$ unique forms. Measuring $C_D$ for every possible genome is beyond the capabilities {even of} our automated setup. This would require a continuous run-time of $\sim 1$ years. To link shape and drag coefficient, we therefore introduce a quantitative shape descriptor, the compactness $Q$
\begin{equation}
	Q = 4\pi \frac{A}{P^2}, \label{eq:Q}
\end{equation}
also known as the isoperimetric quotient \cite{esteban2019study}{. Here, $P$ is the perimeter of the flat lattice animal.} A circular disk ($Q=1$) is the most compact 2D configuration, while $Q\to 0$ for slender geometries (e.g., a thin strip). Mapping the drag coefficient $C_D(A,Q)$ as a function of the area $A$ and compactness $Q$, we hope, will reveal general trends in the shape-performance space. Our data covers a reasonably broad range of compactnesses, $Q=0.13-0.68$, and areas, $A={7}-55$ mm$^2$ (Fig. \ref{fig:fig2}A and \ref{fig:S2}).

\section{Results}
The motion of freely falling objects depends on their size and shape, and like other flying objects, the link is complex and hard to predict \cite{esteban2019study,howison2020large}. To unpack the relationship between form, function, and evolutionary adaptation, we proceed in two steps: First, {31} generic polyominoes comprising biologically inspired lattice animals and randomly generated shapes were selected to determine the link between total area, morphology and resulting drag. These polyominos are shown in Fig. \ref{fig:fig2}  (see also Fig. \ref{fig:S2}). {Secondly}, we explore shape evolution using a genetic algorithm.
%32

\subsection{Size, shape, and aerodynamics performance}
The aerodynamic performance of lattice animals depends on an interplay between size, shape, and flow conditions.
%{We note that the drag coefficients at $Re\approx 10-100$ are complex due to comparable viscous and inertial effects.} 
Examining data from ${31}$ generic shapes reveals the first outlines of a physical picture: The drag coefficient $C_D$ is a decreasing function of both area $A$ and compactness $Q$ (Fig. \ref{fig:fig2}). 
%32

The area dependence is straightforward to understand; larger objects experience higher Reynolds number because it scales with the square root of the area ($Re\sim \sqrt A$, see Eq. \eqref{eq:Re}{, Fig. \ref{fig:S3}}). At the same time, the drag coefficient decreases with $Re$ as the flow transitions from viscous to inertial regimes {\cite{bagheri2016drag}}. Consistent with this idea, the largest drag coefficient observed was that of a small {T-shaped polyomino  ($A= 9$ mm$^2$, $C_D=2.5\pm0.08$, $Re\approx 30$, genome $\mathcal{G}_1=00000\,\ldots\,00001\,00001 (2)$)}, while the smallest $C_D$ was observed for a large rectangular wing ($C_D=1.30\pm0.04$, $A=55$ mm$^2$, $Re\approx {110}$, genome $\mathcal{G}=11111\,\ldots\, 11111 (2)$). 

% wingless polyomino ($\mathcal{G}_1$, $C_D=2.70\pm0.08$, $Re\approx 20$), while the smallest $C_D$ was observed for a large rectangular wing ($\mathcal{G}_4$, $C_D=1.30\pm0.04$, $Re\approx {110}$). 

The experimental link between drag coefficient $C_D$ and  the detailed morphology is not as easy to unpack. However, the differences in geometry can, in some cases, alter the flow sufficiently to substantially influence the settling process and hence the drag coefficient $C_D$ in a predictable manner. For instance, $C_D$ differs by  $\sim 30\%$ between two equal-area forms  (Fig. \ref{fig:fig1}C and \ref{fig:fig2}A). One ($\mathcal G_2$) has a compact oval shape, while the other ($\mathcal G_3$) has a sparse, finger-like appearance. These observation are consistent with previous experiments on snow-like particles \cite{jayaweera1972equivalent,list1971free};  however, the precise link between geometry and drag remains unclear. 

We propose that the close spacing of the fingers restricts the fluid's flow through the gaps, which in turn raises the drag force. This notion is backed by particle tracking observations: fluid flows around the oval lattice animal, and two vortices are observed in the wake,  consistent with observations on disks (Fig. \ref{fig:fig2}(B) and Video S1, \cite{cummins2018separated,willmarth1964steady}). Conversely, fluid does not appear to enter the bristle gaps in the sparse shape (Fig. \ref{fig:fig2}(C) and Video S{1}).  Instead, we observe spanwise flow along the wing edge. This suggests that the gaps between the bristles are not fully permeable. We thus speculate that shear layers form virtual barriers, reducing leakage through the gaps \cite{cheer1987paddles,kolomenskiy2020aerodynamic,jaffar2020leakiness,lee2020optimal}. This effectively slows down the polyomino, causing {an} apparent increase in the drag coefficient $C_D$. 

Data suggest a connection between the genome \(\mathcal{G}\) and drag coefficient \(C_D\), which can be partly explained by fluid-structure interactions. However, detailed shape features that govern the drag coefficient still need further understanding. Notably, the drag coefficient \(C_D \sim \exp(-0.55Q)\) appears to decrease approximately exponentially with the compactness \(Q\) (Fig. \ref{fig:S4}). This supports the idea that rake-like structures impede fluid flow through the gaps, thereby increasing the drag force. In the following section, we will utilize this empirical connection for shape optimization.

%To further explore potential origins of this link between drag and shape, we investigate the wakes of two polyominos: the compact oval $\mathcal G_2$ (compact oval, $Q=0.5$) and the sparse bristled $\mathcal G_3$ ($Q=0.13$). {Flow characterization} reveals that fluid flows around the oval lattice animal $G_2$ (Fig. \ref{fig:fig2}(B) and Video S1). Two vortices are observed in the wake,  consistent with observations on disks  \cite{cummins2018separated,willmarth1964steady}. Surprisingly, fluid does not appear to enter the bristle gaps in the sparse shape $\mathcal G_3$ (Fig. \ref{fig:fig2}(C) and Video S{1}).  Instead, we observe spanwise flow along the wing edge. This suggests that the gaps between the bristles are not fully permeable. We thus speculate that shear layers form virtual barriers, reducing leakage through the gaps \cite{cheer1987paddles,kolomenskiy2020aerodynamic,jaffar2020leakiness,lee2020optimal}. This effectively slows down the polyomino, causing {an} apparent increase in the drag coefficient $C_D$.  {Consequently, the drag coefficient $C_D$ must decrease with the compactness $Q$ as sparser morphologies contain a larger number of appendages between which virtual barriers can form. These outgrowths disappear in compact shapes. 

\subsection{Morphological optimization}
Having explored the basic relations between aerodynamic performance, size, and shape, we now turn our attention to constrained optimization problems. Although we have identified high perfoming morphologies (Fig \ref{fig:fig2}A), it raises the question of whether a unique ideal shape exists for a given constant area $A$. To identify lattice animals with optimal aerodynamic performance, we specifically look for shapes that either maximize or minimize the drag coefficient $C_D$. 

 \begin{figure*}[t!]
    \centering
    \includegraphics[width = 18.5cm]{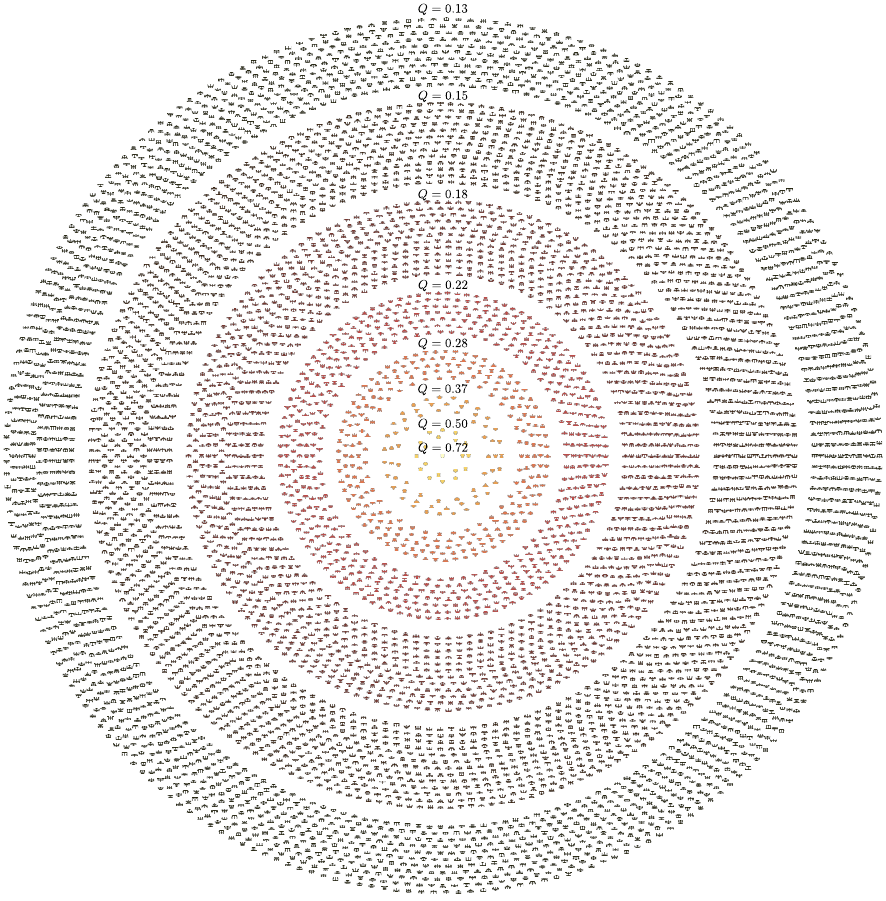}
    \caption{{Taxonomy of lattice animals. We identified $7325$ unique polyominoes of area $A=23$ mm$^2$. Here, they are grouped by compactness $Q$ (Eq. \eqref{eq:Q}). Sparse morpholgies (small $Q$) are abundant, while compact polyominos (large $Q$) are relatively rare. Evolutionary algorithms, which rely on random mutations, will therefore spend most time in the abundant low-$Q$ states.}}
    \label{fig:uniqueInsects}
\end{figure*}

Performing an exhaustive {search} among the $\sim 10^4$ unique forms is intractable {(Fig. \ref{fig:uniqueInsects})} because it would require a continuous run-time of one year. Therefore, we position our insect model in an evolutionary context by relinquishing control of the morphologies to a genetic algorithm. {Shape} information (i.e., genes) are transferred from one generation to the next in proportion to aerodynamic performance. Our implementation follows \citet{ramananarivo2019improving}, and we expect that preferential selection and breeding will lead to progressive performance improvement \cite{spall2005introduction,ramananarivo2019improving}. Note the novel feature of our approach: because the ASAP setup is fully automated, the optimization scheme can run unsupervised\footnote{Accumulating paper shapes at the bottom are sifted out every 3-4 generations. Intervention required in case of critical software or hardware failure.}. The experiments are performed at fixed area $A=23$ mm$^2$. See additional details in the supplemental materials. 

\begin{figure}[t!]
    \centering
    \includegraphics[width = 8cm]{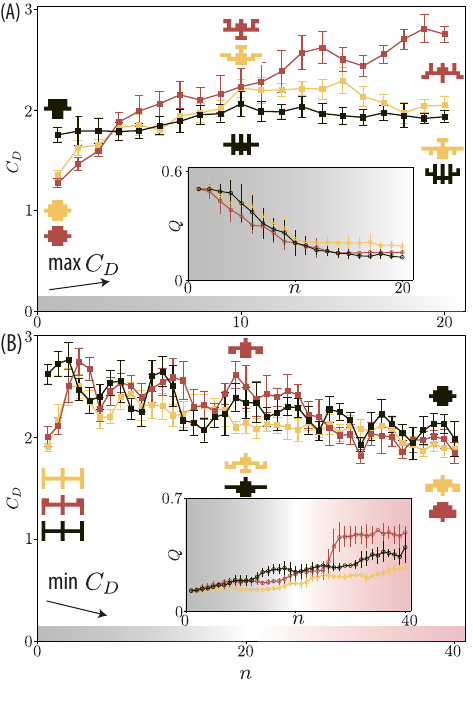}
    \caption{Artificial evolution experiments. (A) The drag coefficient $C_D$ is plotted against generation $n$ for a genetic experiment aimed at maximizing $C_D$.  
(B) The drag coefficient $C_D$ is plotted against generation $n$ for a genetic experiment focused on minimizing $C_D$.  
We note that the forward evolutionary process (A) is faster  compared to the backward one (B).  
Inserts in (A) and (B) show the evolution of compactness $Q$ (Eq. \eqref{eq:Q}). Error bars indicate population averages (data presented in Fig. \ref{fig:S5}, \ref{fig:S6}, and \ref{fig:S7}). The experiments were conducted with a fixed area of $A=23$ mm$^2$ and a population size of $N=10$. For more information, see the details of the genetic algorithm in the supplemental materials.
  }
    \label{fig:fig3}
\end{figure}

We begin by considering the case of maximizing the drag coefficient $C_D$. This corresponds to the fitness function $f=C_D-C_{D,\mathrm{min}}$, which ensures that the poorest performing morphology in each generation {always} has the lowest fitness ($f=0$). It therefore never passes on genetic information. In contrast, the best shape (fitness $f=C_{D,\text{max}}-C_{D,\text{min}}$) has the highest propability of producing offspring. 

Starting near a compact shape (\(\mathcal{G}_{2}=4575\,(10)\), \(Q=0.50\)), the drag coefficient \(C_D\) steadily increases over \(n=20\) generations (Fig. \ref{fig:fig3}A). Bristle-like wings evolve; however, the emerging populations are not replicated when the evolutionary experiment is repeated. While unique forms rarely recur, the average compactness \(Q\) clearly increases from generation to generation (Fig. \ref{fig:fig3}A (insert)). This trend aligns with the idea of a link between compactness \(Q\) and drag coefficient \(C_D\) (Fig. \ref{fig:S4}).  

Another notable feature of high-performance geometries is that the finger-like structures are roughly evenly spaced. To understand the specific distance between appendages (approximately \(1\) mm), we examine a small hole (radius \(r\)) in a planar object settling at a speed of about \(1\) cm/s. The pore becomes invisible to the fluid if the pressure-driven flow through the hole moves at speeds \(v\) slower than the falling velocity \(U_z\). But how small must the hole be? We can estimate the flow speed \(v\) in the pore using Sampson's solution to Stokes' equation, which gives \(v = {r \Delta p}/{(3\pi \eta)}\). Here, the pressure drop \(\Delta p = \Delta \rho gt\) is caused by the weight of the settling object. The limiting pore size (corresponding to \(v=U_z\)) is \(r = {3\pi \eta U_z}/{(gt\Delta \rho)} \sim 1\, \mathrm{mm}\), which is consistent with the observed spacing of the appendages.

%However, it also implies a relatively flat fitness landscape in which many genetic variants have nearly identical fitness because they correspond to the same compactness.

Reversing the direction of evolution reveals other interesting {features} of the problem (Fig. \ref{fig:fig3}B). To minimize $C_D$, we consider the fitness function $f=C_{D,\mathrm{max}}-C_{D}$, and begin our search starting from near relatively sparse morphology ($\mathcal{G}_{3}=32641156\,(10), Q=0.13$). Again, the population trajectories are not reproducible, but the shift towards more compact individuals is clear (Fig. \ref{fig:fig3}B, insert). 

Notably, the evolution towards minimum $C_D$ proceeds significantly slower than in the direction of maximum $C_D$. {Our data indicates that the process is at least 50\% slower. While a firm understanding of the hysteresis remains unknown, we present a potential explanation in the following section.}

\subsection{Free fitness dictates the rate of evolution}
The artificial evolution experiments revealed a dependence on directionality (Fig. \ref{fig:fig3}). Evolving slowly settling polyominoes occurs about twice as fast as rapidly settling geometries. This is surprising because the numerical fitness gain is the same in both experiments. A natural question arises: Why does the adaptation rate depend on the direction of evolution?

Examining the morphological space (Fig. \ref{fig:uniqueInsects}) may provide clues for a possible explanation: there are many more sparse morphologies than compact geometries. Since sparse shapes typically have higher drag, they are presumably easier for the evolutionary process to identify. Hence this is a faster process. To formalise this idea, we assess the genetic algorithm results in terms of a free fitness function, which captures both the applied evolutionary pressure and entropic effects \cite{iwasa1988free,sella2005application,manrubia2021genotypes}. We define the free fitness as
\begin{equation}
	F=\frac{f}{f_0}+\kappa \frac{S}{S_0} \label{eq:freefitness}.
\end{equation}
where $f_0$ and $S_0$ is the fitness and entropy of a reference state ($\mathcal G_3$ with $C_D=2.19$ and $Q=0.13$, see Fig. \ref{fig:fig1} and \ref{fig:fig2}). The parameter $\kappa$ is the evolutionary temperature \cite{sella2005application} {which can vary with, e.g., the population size \cite{sella2005application}, the mutation rate \cite{manrubia2021genotypes}, or the experimental variability (Fig. \ref{fig:fig2}A, inserts). For simplicity, we set  $\kappa =1$. This consistent with a monotonically increasing free fitness, $F$.} Please also note that in contrast to a free energy, the free fitness $F$ always increases {\cite{iwasa1988free,manrubia2021genotypes,sella2005application}}).

\begin{figure}[t!]
    \centering
    \includegraphics[width = 8cm]{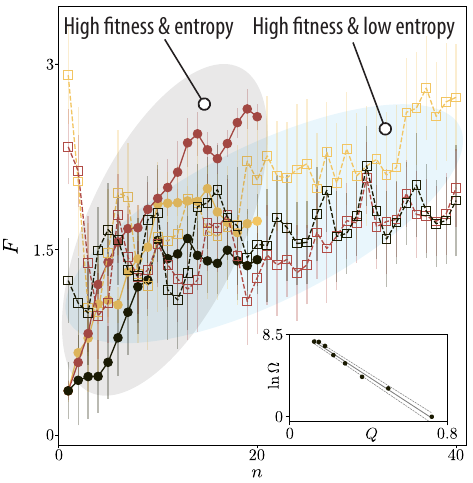}
    \caption{Free fitness determines the evolution rate. The free fitness $F$ (Eq. \ref{eq:freefitness}) is depicted as a function of generation $n$, showing an increase with each generation. However, experiments aimed at achieving maximum drag coefficients (solid dots) quickly progress towards states that exhibit both high fitness and significant entropy. This acceleration occurs due to the abundance of {geometries} associated with high $C_D$ (or low compactness $Q$). In contrast, data pursuing minimum $C_D$ (high $Q$) must carefully balance high fitness against lower entropy. This lays the foundation for understanding the evolutionary hysteresis observed in the genetic data (Fig. \ref{fig:fig3}A,B). The inset illustrates the number of distinct $5\times 5$ bingo-plate morphologies $	\ln {\Omega }$ as a function of compactness $Q$. }
    \label{fig:fig4}
\end{figure}

To unpack the physical meaning of the the entropic term, we note that each performance value, i.e., drag coefficient $C_D$, is shared by a (potentially large) number $\Omega (C_D)$ of individuals. It is therefore more likely to randomly select some $C_D$ values than others. With the entropy $S=\ln (\Omega)$, we can estimate the weighting by exploiting the experimental link between drag $C_D$ and compactness $Q$ (Fig. \ref{fig:fig2}{, Fig. S4}). Expressing the entropy $S$ in terms of the compactness $Q$ leads to a particularly simple result: The entropy is a linear function of compactness $S=9.5 -13.3Q$ (Fig. \ref{fig:fig4}, {insert}). We note that the constants are specific to this problem, and cannot be applied outside the $5\times 5$ bingo-plate wing geometry.

Introducing the free fitness allows us to evaluate the evolutionary dynamics of emerging polyominoes (Fig. \ref{fig:fig4}). During evolution towards maximum $C_D$, both the fitness and entropic term in Eq. \ref{eq:freefitness}, increase between generations. This speeds up the process, because {many geometries have} large $C_D$ (or low compactness $Q$). In contrast, entropy must decrease to minimize $C_D$ (maximize $Q$). This reduces the net rate of fitness improvement between generations \cite{sella2005application,manrubia2021genotypes}.  We speculate that the lack of convergence in, e.g., insect wing shape, may, in part, reflect the competition between fitness and entropy, making it difficult to evolve a truly optimal form.

\section{Discussion and conclusion}
A clearer picture of the relationship between the shape and aerodynamic performance of lattice animal wings has emerged. Data from our experiments (Fig. \ref{fig:fig2}) indicate that for a fixed area (mass), the drag coefficient $C_D$ (Eq. \eqref{eq:drag}) correlates negatively with the compactness $Q$ (Eq. \eqref{eq:Q}). Bristled morphologies (i.e., low $Q$) are effective flyers because viscous drag limits flow between the lobes, which increases the drag force. This supports the idea that bristled insect wings are effective fliers \cite{sun2023colloquium}.

Another possible reason for the existence of complex wing forms stems from our artificial evolution experiments (Fig. \ref{fig:fig3} and \ref{fig:fig4}). Simple shapes, such as circles and rectangles, are relatively rare, while there are many more ways of generating complex and sparse geometries (Fig. \ref{fig:uniqueInsects}). Mutations therefore cause evolutionary algorithms to spend more time in these high-entropy states. While this does not explain the coexistence of bristled and membranous wings, it offers a complementary perspective on the emergence of complex wing morphologies.

%\nocite{*}

%\bibliography{bib}% Produces the bibliography via BibTeX.

%%%bbl file paste
\providecommand{\noopsort}[1]{}\providecommand{\singleletter}[1]{#1}%
%

%%% end bbl file paste

\newpage
\clearpage
\newpage

%\onecolumn
\onecolumngrid
\setcounter{page}{1} 
%\subsection*{Arnbjerg-Nielsen \emph{et al.}}
%\maketitle
\section{Supplementary materials}
\label{sec:som}
\renewcommand{\thefigure}{S\arabic{figure}}
\setcounter{figure}{0}
%\section{Materials and Methods}
\subsection{Experimental setup}
The terminal velocity of the polyominoes is experimentally measured with both shape cutout and free fall tracking performed inside a closed loop (see, e.g. \citet{howison2020large}). We use a solution of soap dish (Shine Classic Washing-Up Liquid, Mayeri Industries, Estonia) and disinfectant (Cif Professional 2in1 Cleaner Disinfectant, Pro Formula, Denmark) to ensure the paper shape breaches the water-air interface. The camera is attached to a Jetson Nano 4GB (Nvidia, USA), where tracking is performed in real-time with an OpenCV contour finding algorithm and CUDA accelerated background subtraction. Everything is automated over Python (version 3.8.10 or 3.9.15, Python, USA). Accumulating paper shapes are fished out approximately every third or fourth generation with a subsequent 10-minute break to reestablish quiescent water for the next measurements. In case of critical software or hardware failure, the generation is restarted by rerunning the genetic algorithm at the current generation and starting the hardware loop.

\begin{figure}[htb]%
\centering
\includegraphics[width=8cm]{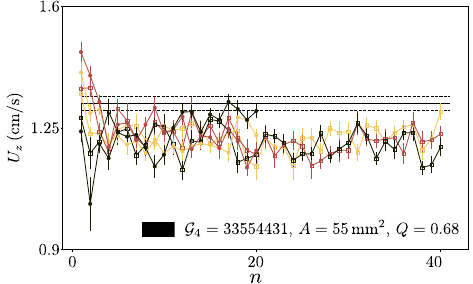}
\caption{Measured terminal velocity for the largest morphology, $\mathcal{G}_{4}=33554431$ at each generational step. Data both maximizing (solid circles) or minimizing (open squares) drag is included. The solid line is the value measured for the initial sample with the standard deviation of the mean marked with a dashed line. {5,056 experiments are included in this data set.}}
\label{fig:backgroundVariation}
\end{figure}

Fluctuations in the measured terminal velocity, $U_z$, are observed by implementing a reference shape, $\mathcal{G}_4=33554431$, in all populations (Fig. \ref{fig:backgroundVariation}). These are contributed to the sensitivity in environmental and initial conditions \cite{tinklenberg2023thin,howison2020large,esteban2019study}, such as, e.g., dissolving paper and fluctuating temperature. Randomization of shape sequence and orientation is introduced to avoid bias within a each population.

\subsection*{Genetic algorithm and breeding function}
The drag performance of insect inspired polyominoes are iteratively improved by implementing a genetic algorithm with population size $N=10$, roulette-wheel-parent selection, single bit mutation \cite{lawrence1991handbook} with 20\% probability, and generational replacement with transfer of the best-performing morphology to the succeeding generation adapted from \citet{ramananarivo2019improving}. %First, the evolutionary algorithm is described followed by the breeding procedure.

The first generation is homogenously populated with an initial shape ($\mathcal{G}_{2}=4575$ for maximizing drag, $\mathcal{G}_{3}=32641156$ when minimizing drag), each mutated with a probability of 20\%. Measuring their drag coefficient, $C_D$, their fitness is evaluated and ranked according to the fitness function, 
\resizebox{\linewidth}{!}{\parbox{1.15\linewidth}{%
\begin{subnumcases}{f(C_{D})=}
			C_{D} -C_{D,min} & towards high drag, \label{eq:forwardsevolution}
			\\
			C_{D,max}-C_{D} & towards low drag. \label{eq:backwardsevolution}
\end{subnumcases}
}}

Here $C_{D,min}$ is the lowest and $C_{D,max}$ is the highest drag measured within the population. $C_{D}$ is the drag for each individual. This linearly ranks the performance with $f=0$ for the lowest-performing shape.

The best-performing morphology is carried over unmodified with full replacement of the rest of the population. The next generation is populated by preferentially cross-breeding morphologies from the current generation, repeating the process nine times (custom Python script, version 3.8.10 or 3.9.15, Python, USA).  Two parents, ($\mathbf{P_1}$ and $\mathbf{P_2}$), are independently selected (allowing self-breeding), with probability of selection proportinate to the fitness
\begin{equation}
	p_i = \frac{f_i}{\sum f_i}.
\end{equation}

\begin{figure}[htb]%
\centering
\includegraphics[width=8cm]{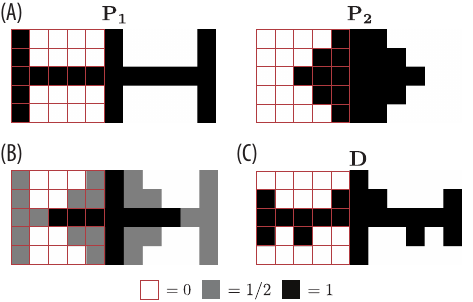}
\caption{Recombination of two parents into a new daughter. The wing-spanning matrix is marked with a red grid. (A) Two parents. (B) Computed average. Overlapping elements, $x_{i,j}=1$ (black), are kept, and the remaining elements are sampled randomly where only one parent has mass, $x_{i,j}=0.5$ (grey). (C) Daughter after recombination and mutation.}\label{fig:figBreeding}
\end{figure}

\begin{figure}[h]%
\centering
\includegraphics[width=15cm]{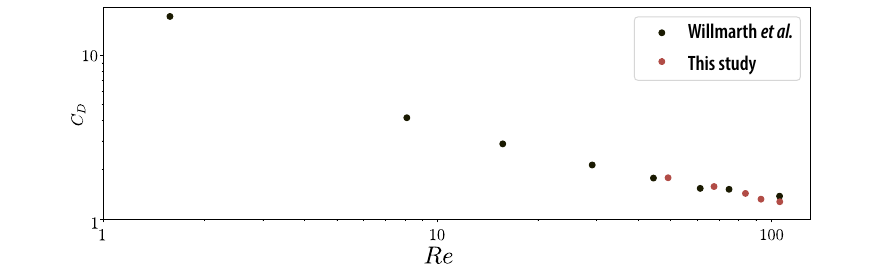}
\caption{Drag coefficient $C_D$ plotted as a function of Reynolds-number $Re$ for circular disks. Data (red dots) are consistent with previous observations (black dots) by \citet{willmarth1964steady}. The errorbars are smaller than plot markers.\label{fig:S1}}
\end{figure}

\begin{figure}[h]%
\centering
\includegraphics[width=\textwidth]{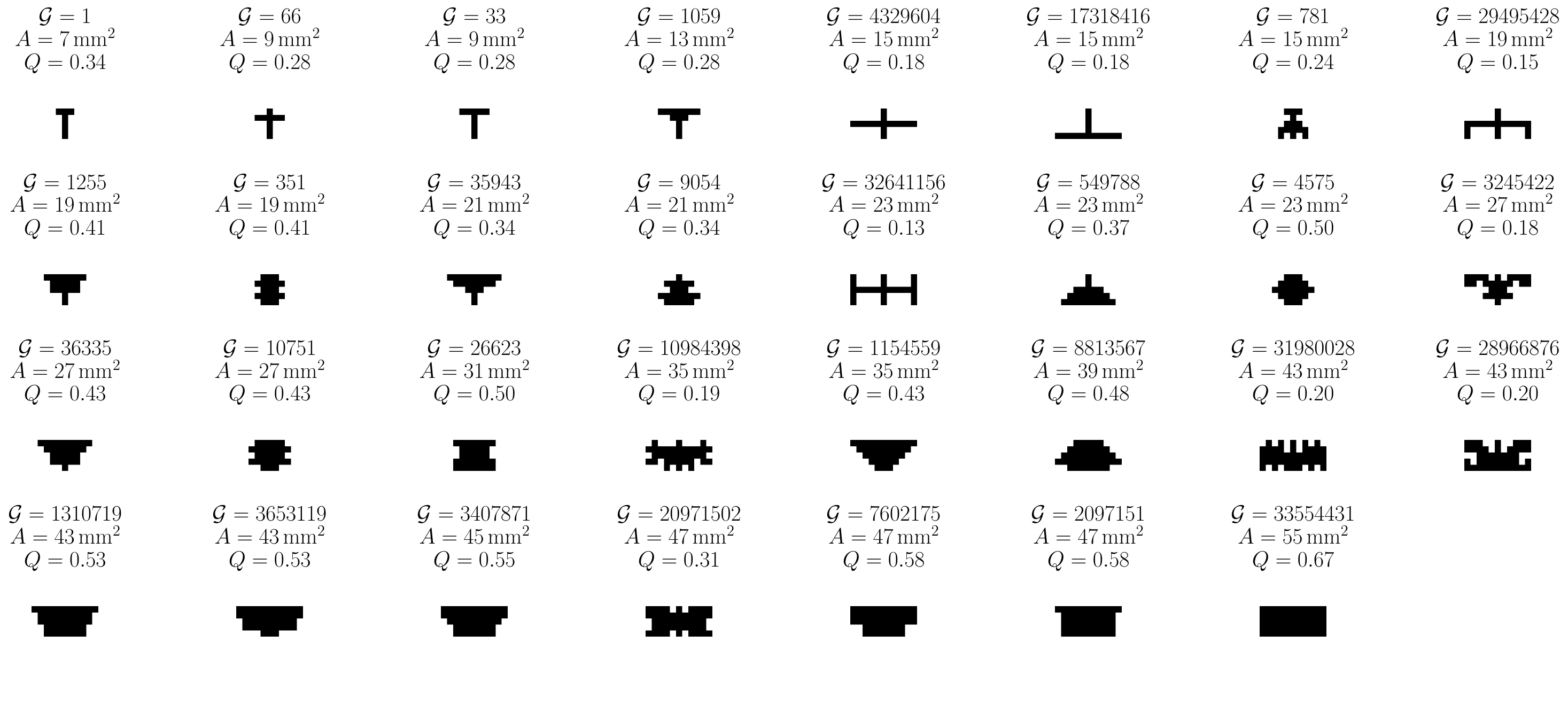}
\caption{Randomly generated and biomimicking lattice animals used to determine relation between size, shape, and hydrodynamic performance. %Each polyomino is identified by concatenating its "wing" matrix columns and converting the resulting binary number to a decimal representation. 
The genome $\mathcal G$, area $A$, and compactness, $Q=4\pi A/P^2$ identify each polyomino.\label{fig:S2}}
\end{figure}

\begin{figure}[h]%
\centering
\includegraphics[width=15cm]{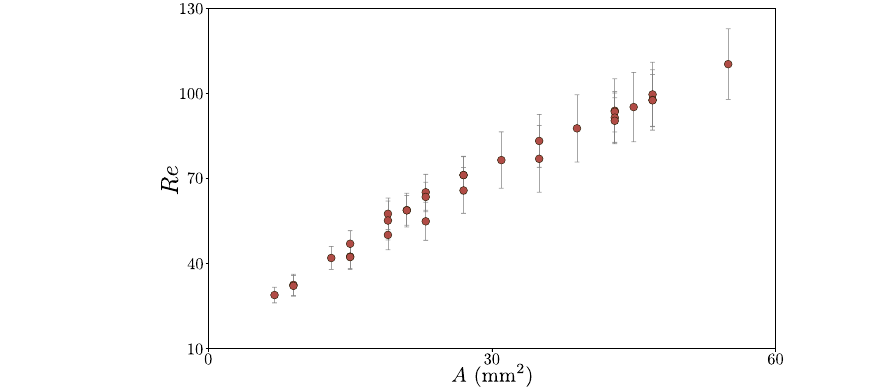}
\caption{Measured Reynolds number $Re$ plotted as a function of total area $A$ for the sample set shown in Figure S{2}. The Reynolds number (Eq. \eqref{eq:Re}) scales with the square root of the area $A$. Note that the detailed shape (i.e., genome $\mathcal{G}$) also influences the terminal velocity, and hence $Re$. Hence two equal area forms can differ slightly in observed $Re$.\label{fig:S3}}
\end{figure}

\begin{figure}[h]%
\centering
\includegraphics[width=15cm]{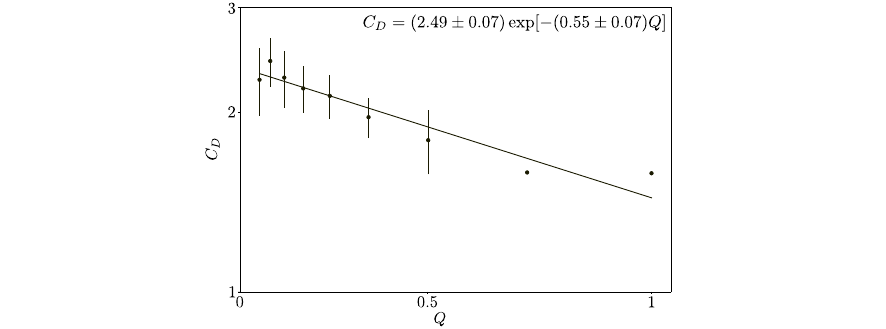}
\caption{The relationship between drag coefficient $C_D$ and compactness $Q$ is examined through experiments conducted at a constant area of $A=23$ mm$^2$ (data from Fig. \ref{fig:fig2} and \ref{fig:fig3}). A least squares fit is consistent with an exponential decay. Note that the $Q=1$ data point corresponds to a circular disk of $A=25$ mm$^2$. \label{fig:S4}}
\end{figure}

Recombining the parents to a daughter ($\mathbf{D}$) is based on the average of their wing-spanning matrix (Fig. \ref{fig:figBreeding}). Selecting nine elements to ensure area $A=23$ mm$^2$, overlapping matrix elements (value $1$) are always chosen, with the remaining elements randomly elected from non-overlapping elements (value $1/2$). The daughter must be simply rook-wise connected \cite{read1962contributions,golomb1996polyominoes} before subjecting it to a bit mutation. While the final lattice animal bears resemblance to both parents, new characteristics can also present themselves (Fig. \ref{fig:figBreeding}). The source code for the breeding algorithm can be found at our \href{https://github.com/Jensen-Lab/PolyominoWingEvolution/blob/4caa0845b7e0f55f4288a9790a9b7a38a94224d4/insectGeneration.ipynb}{GitHub repository} \cite{githubBreeding}.

\begin{sidewaysfigure}[ht]
    \includegraphics[width=\textwidth]{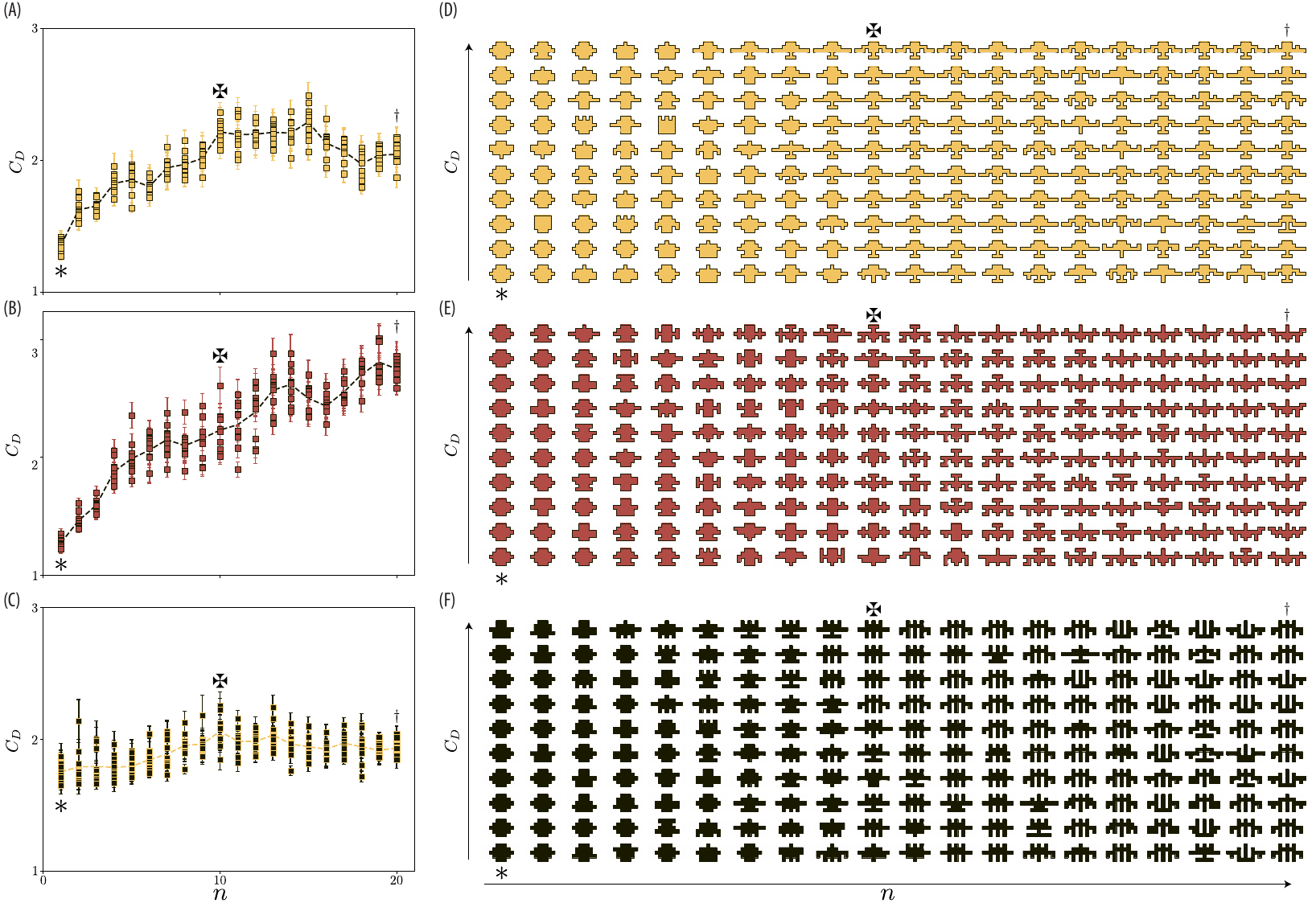}
    \caption{Detailed data from genetic algorithm with selection pressure towards high drag (Fig. \ref{fig:fig3}A, fitness function, Eq. \eqref{eq:forwardsevolution}) applied to lattice animals at constant area, $A=23$ mm$^2$.  Each measurement of the drag coefficient, $C_D$ (squares), and the compactness, $Q$ (circles) are plotted along with the average (dashed black line) (A-C). The corresponding population is drawn (D-F). The worst performing (\textasteriskcentered, lowest $C_D$) in the first generation, best performing half-way (\maltese, highest $C_D$) and after ended algorithm (\dag, highest $C_D$) are marked. \label{fig:S5}}
\end{sidewaysfigure}

\begin{figure}[ht]
    \includegraphics[width=0.9\textwidth]{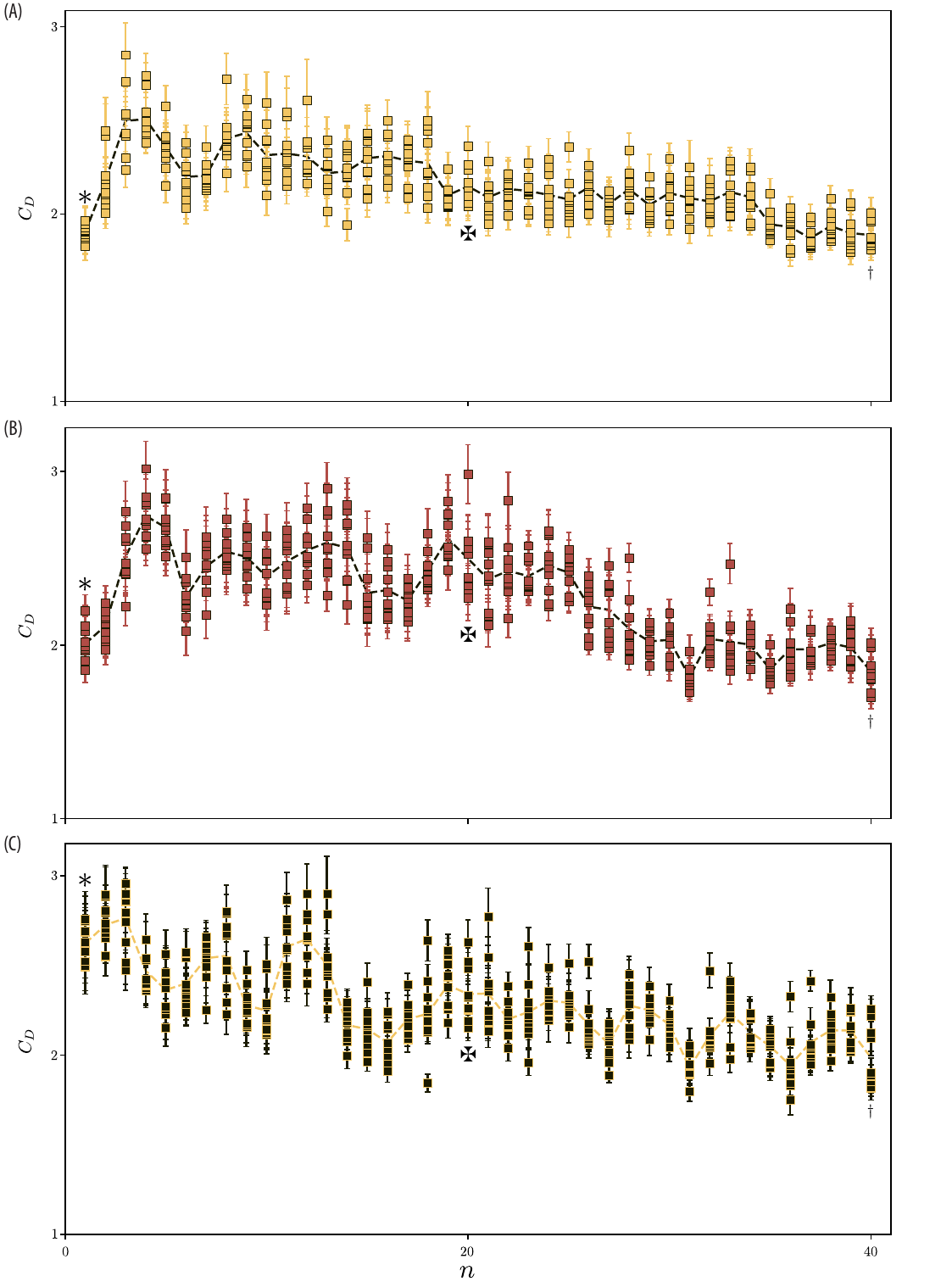}
        \caption{Detailed data from genetic algorithm with selection pressure towards low drag (Fig. \ref{fig:fig3}B, fitness function, Eq. \eqref{eq:backwardsevolution}) applied to lattice animals at constant area, $A=23$ mm$^2$.  Each measurement of the drag coefficient, $C_D$ (squares), and the compactness, $Q$ (circles) are plotted along with the average (dashed black line) (A-C). The corresponding population is drawn in Fig. \ref{fig:S7}. The worst performing (\textasteriskcentered, highest $C_D$) in the first generation, best performing half-way (\maltese, lowest $C_D$) and after ended algorithm (\dag, lowest $C_D$) are marked. \label{fig:S6}}
\end{figure}

\begin{sidewaysfigure}[ht]
    \includegraphics[width=\textwidth]{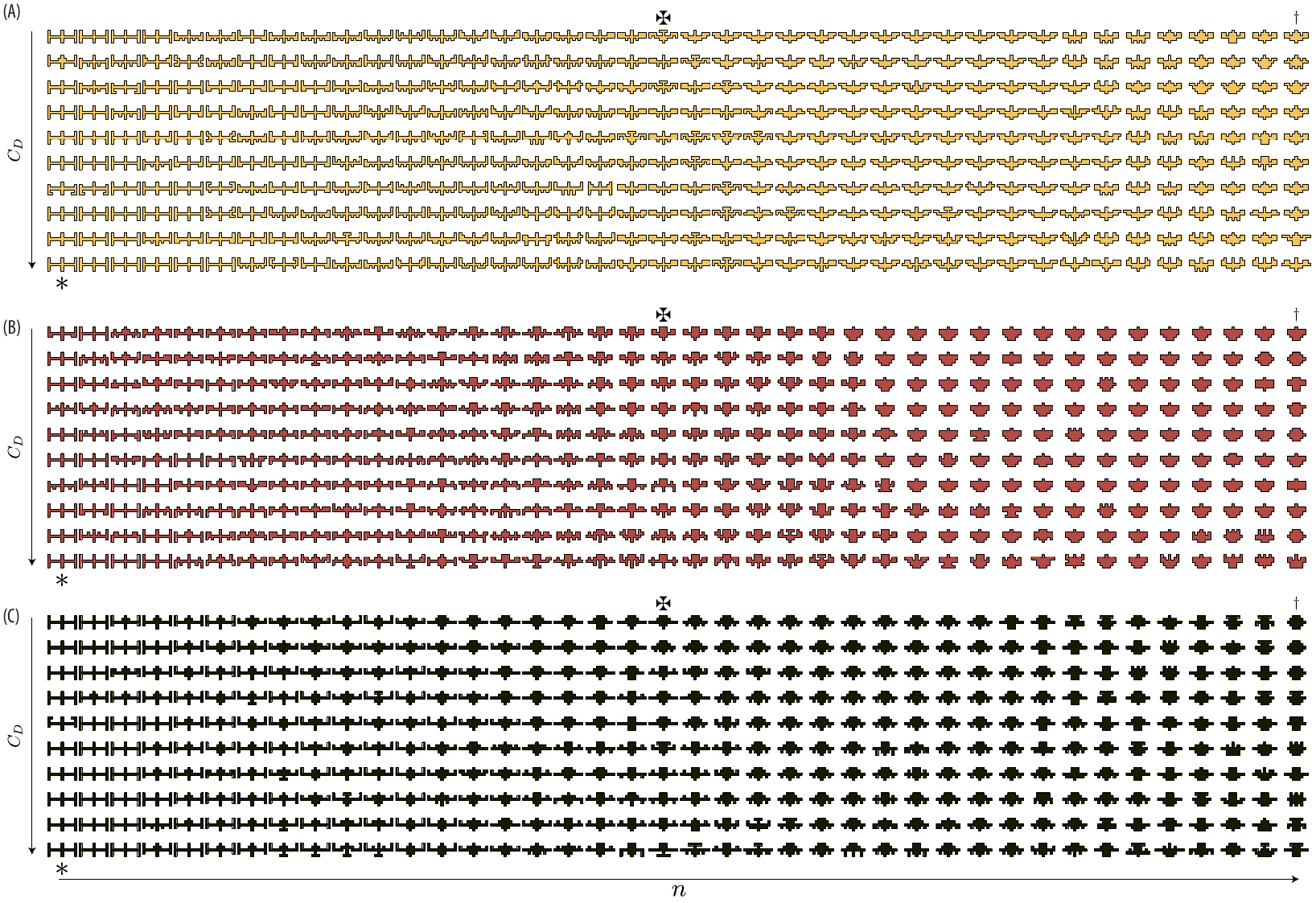}
        \caption{Detailed data from genetic algorithm with selection pressure towards low drag (Fig. \eqref{fig:fig3}B, fitness function, Eq. \ref{eq:backwardsevolution}) applied to lattice animals at constant area, $A=23$ mm$^2$.  The detailed populations are drawn (A-C). Each measurement of the drag coefficient, $C_D$ (squares), and the compactness, $Q$ (circles) are plotted along with the average (dashed black line) in Fig. \ref{fig:S6}. The worst performing (\textasteriskcentered, highest $C_D$) in the first generation, best performing half-way (\maltese, lowest $C_D$) and after ended algorithm (\dag, lowest $C_D$) are marked. \label{fig:S7}}
\end{sidewaysfigure}

\begin{figure}[h]%
\centering
\includegraphics[width=5cm]{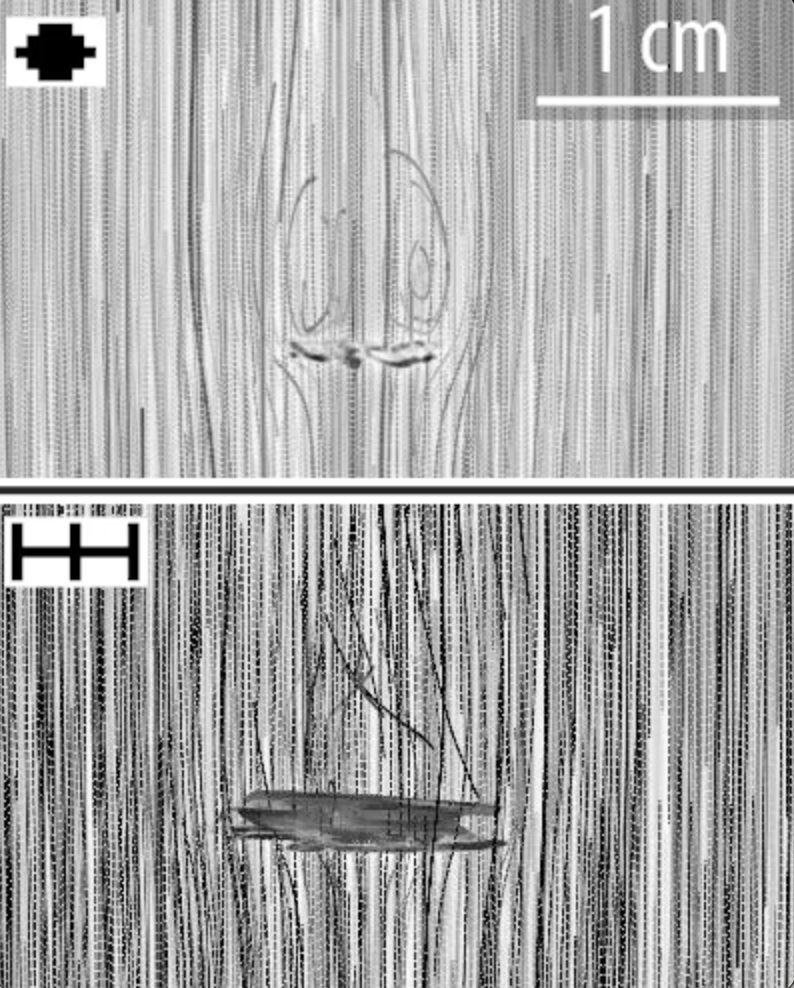}
\caption{VIDEO S1: Representative streamlines for settling dynamics of lattice animals with genomes $\mathcal{G}_{2}=4575$ and $\mathcal{G}_{3}=32641156$. Streamlines are obtained using Flowtrace for ImageJ / Fiji \cite{gilpin2017flowtrace}, merged over 300 images with setting \textit{substract median} enabled. Video is played at 50\% speed: 50 fps, data obtained with 100 fps.\label{fig:V1}}
\end{figure}

\end{document}